# The confinement energy of quantum dots

Samrat Dey*[1], Devkant Swargiary[2], kishan Chakraborty[2], Debasmita Dasgupta[2], Darsana Bordoloi[2], Rituja Saikia[2], Darsana Neog[2], Shishila Shimray[2], Supriyanka Paul[2], Kabita Brahma[2], Joydeep Dey[2] and Saurav Choudhury[2]

[1]Department of Physics, DBCET, Assam Don Bosco University,
Guwahati -781017, INDIA.

[2]Department of Electronics and Communication Engineering, DBCET, Assam Don Bosco University,
Guwahati -781017, INDIA.

Abstract

One of the most significant research interests in the field of electronics is that on quantum dot, because such materials have electronic properties intermediate between those of bulk semiconductors and those of discrete molecules. Confinement energy is a very important property of quantum dot. In this study, quantum confinement energy of a quantum dot is concluded to be $h^2/8md^2$ (*d* being the diameter of the confinement) and not $h^2/8ma^2$ (*a* being the radius of the confinement), as reported in the available literature. This is in the light of a recent study [1]. This finding should have a significant impact in the understanding of the physics of quantum dot and its technological application.

---

* Corresponding Author

## 1. Introduction

A quantum dot is a portion of matter (e.g., semiconductor) whose excitons are confined in all three spatial dimensions[2]. They exhibit size dependent optical and electrical properties. Apart from having scope in optical applications and applications in quantum computer, in electronics they can operate like a single electron transistor and show the Coulomb blockade effect[2, 3]. Quantum dots are thus of significant research interest from the point of view of electronics.

According to literature[2] the total energy of quantum dot is

$$E = E_{band\ gap} + E_{confinement} + E_{excitation} \qquad (i)$$

$E_{confinement}$ can be calculated by considering the model of a particle trapped in a cavity of radius $a$ which according to standard texts [4-6] is

$$E_{confinement} = h^2/8ma^2 \qquad (ii)$$

However, according to a recent study [1] we suggest that

$$E_{confinement} = h^2/8md^2 \qquad (iii)$$

where $d$ is the diameter of the confinement. Let us analyze the result in detail.

## 2. Analysis

It is discussed in [1] that the use of conventional approaches of quantum mechanics in obtaining Eqn. (ii) (as concluded in [4-6]) is not proper. The ground state energy of a particle in a spherically symmetric confinement is typically found by solving the radial part of the Schrodinger equation (for, $V = 0$)

$$\frac{1}{r^2}\frac{d}{dr}\left(r^2 \frac{dR_l(r)}{dr}\right) + \left[\frac{2ME}{\hbar^2} - \frac{l(l+1)}{r^2}\right] R_l(r) = 0, \qquad (iv)$$

where, $M$ = mass of the particle, $\hbar = h/2$ (with $h$ being the Planck's constant) and $R_l(r)$ (with $l$ representing orbital angular momentum of the state of the particle) is an eigen function of eigen energy $E$, and applying the boundary condition that $R_l(r) = 0$, for $r = a$ (where $V$ tends to infinity). The value of $R_0(r)$ and $E_0$ (ground state energy eigen value) so obtained are, respectively,

$$R_0(r) = B\frac{\sin kr}{r} \qquad (v)$$

and

$$E_0 = \frac{h^2}{8ma^2}, \qquad (vi)$$

$B$ being the normalization constant. However, as discussed in [1], in doing so an implicit hard core potential at the centre of the cavity comes into the picture, although physically it is not present (i.e., although no boundary condition is there at $r = 0$, according to the physical description of the system). This results in a forced node in the wave function at that point, although physically there is no potential to compel that node. In fact, had there been such a point hard core potential present at the centre, the confinement energy eigen value would have surely been given by Eqn. (ii). However, in absence of such a potential the correct confinement energy of such a trapped particle should not be given by Eqn. (ii).

In simple words, a particle /exciton will never encounter a hard core potential at the centre of the cavity, and so it shall not have a forced node at that point. In absence of that forced node the de Broglie wave wavelength for ground state should be longer (by a factor of 2) than what is conventionally understood; consequently, the energy should be lesser (by a factor of 4). Thus, we can show that the true ground state energy of a particle in confinement should be given by Eqn. (iii).

In the light of above discussion, we must conclude that the confinement energy of an exciton trapped in a cavity is given by Eqn. (iii) and not by Eqn. (ii). Thus the confinement energy should be lesser than what is reported in the text. We predict that careful experiments on quantum dots should confirm it.

### 3. Conclusions

Physics of quantum dots has a tremendous scope in nano-electronics which may revolutionise technology. However, correct understanding of energy of quantum dots is necessary for it, specially the quantum confinement energy. This article concludes the correct confinement energy of quantum dots in the light of [1] which is calculated to be lesser than what is available in the standard literatures. We also suggest that careful experiments on quantum dots shall reveal that the confinement energy is much lesser than what is believed till now.

*Acknowledgment*: The authors are thankful to Prof. Yatendra S. Jain, Mr. Sukumar Dey and Dr. Sunandan Baruah for useful interaction.